\documentclass[daat]{ipart}
\usepackage{float}
\RequirePackage{hyperref}

\endlocaldefs

\usepackage{listings}
\usepackage{xcolor}

\def \tma {\textcolor{magenta}}

\usepackage{tikz}
\usetikzlibrary{decorations.pathreplacing}

\pubyear{2025}
\volume{1}
\issue{1}
\firstpage{1}
\lastpage{8}
\arxiv{2506.08976}

\begin{document}

\begin{frontmatter}
\title[Yau-YauAL: A computer tool for solving nonlinear filtering problems]{Yau-YauAL: A computer tool for solving nonlinear filtering problems
\protect
% \thanksref{T1}
}
% \thankstext{T1}{Footnote to the title with the `thankstext' command.}

\begin{aug}
    \author{\inits{Y.}\fnms{Yu} 
    \snm{Wang}\ead[label=e1]{yuwang@bimsa.cn}},
    \address{Beijing Key Laboratory of Topological          Statistics and Applications for Complex Systems, Beijing Institute of Mathematical Sciences and Applications\\
    Beijing 101408, P.R. China\\
    Institute of Statistics and Big Data, Renmin University of China\\
    Beijing 100872, P.R. China\\
    \printead{e1}}  
    \author{\inits{S.}\fnms{Shuyuan} 
    \snm{Xu}\ead[label=e2]{xushuyuan@bimsa.cn}},
    \address{Beijing Key Laboratory of Topological          Statistics and Applications for Complex Systems, Beijing Institute of Mathematical Sciences and Applications\\
    Beijing 101408, P.R. China\\
    Academy of Mathematics and Systems Science, Chinese Academy of Sciences\\
    Beijing 100190, China\\
    University of Chinese Academy of Sciences\\
    Beijing 100049, P.R. China\\
    \printead{e2}}
    \author{\inits{X.}\fnms{Xueda} 
    \snm{Wei}\ead[label=e3]{weixueda@bimsa.cn}},
    \address{Beijing Key Laboratory of Topological          Statistics and Applications for Complex Systems, Beijing Institute of Mathematical Sciences and Applications\\
    Beijing 101408, P.R. China\\
    Institute of Statistics and Big Data, Renmin University of China\\
    Beijing 100872, P.R. China\\
    \printead{e3}}
    \author{\inits{X.}\fnms{Xinrui} 
    \snm{Luo}\ead[label=e4]{luoxinrui@bimsa.cn}},
    \address{College of Mathematics Science, Inner Mongolia Normal University\\
    Hohhot 010022, P.R. China\\
    Beijing Key Laboratory of Topological          Statistics and Applications for Complex Systems, Beijing Institute of Mathematical Sciences and Applications\\
    Beijing 101408, P.R. China\\
    \printead{e4}}
    \author{\inits{S.}\fnms{Stephen} 
    \snm{Shing-Toung Yau}\ead[label=e5]{yau@uic.edu}},
    \address{Beijing Key Laboratory of Topological Statistics and Applications for Complex Systems, Beijing Institute of Mathematical Sciences and Applications\\
    Beijing 101408, P.R. China\\
    Department of Mathematical Sciences, Tsinghua University\\ Beijing 100084, P. R. China\\
    \printead{e5}}
    \author{\inits{S.}\fnms{Shing-Tung} 
    \snm{Yau}\ead[label=e6]{styau@tsinghua.edu.cn}},
    \address{Beijing Key Laboratory of Topological Statistics and Applications for Complex Systems, Beijing Institute of Mathematical Sciences and Applications\\
    Beijing 101408, P.R. China\\
    Yau Mathematical Sciences Center, Tsinghua University\\
    Beijing 100084, P.R. China\\
    \printead{e6}}
    \and
    \author{\inits{R.}\fnms{Rongling} 
    \snm{Wu}
    \thanksref{t2}
    \ead[label=e7]{ronglingwu@bimsa.cn}}
    \address{Beijing Key Laboratory of Topological Statistics and Applications for Complex Systems, Beijing Institute of Mathematical Sciences and Applications\\
    Beijing 101408, P.R. China\\
    Yau Mathematical Sciences Center, Tsinghua University\\
    Beijing 100084, P.R. China\\
    \printead{e7}}
    \thankstext{t2}{Corresponding author.}
\end{aug}
\received{\sday{11} \smonth{6} \syear{2025}}

\begin{abstract}
The Yau-Yau nonlinear filter has increasingly emerged as a powerful tool to study stochastic complex systems. To leverage it to a wider spectrum of application scenarios, we pack the Yau-Yau filtering ALgorithms (YauYauAL) into a package of computer software. Yau-YauAL was written in R, designed to simplify the implementation of the Yau-Yau filter for solving nonlinear filtering problems. Combining R’s accessibility with C++ (via Rcpp) for computational efficiency, YauYauAL provides an intuitive Shiny-based interface that enables real-time parameter adjustment and result visualization. At its core, YauYauAL employs finite difference methods to numerically solve the Kolmogorov forward equation, ensuring a stable and accurate solution even for complex systems. YauYauAL’s modular design and open-source framework further encourage customization and community-driven development. YauYauAL aims to bridge the gap between theoretical nonlinear filtering methods and practical applications, without requiring expertise in differential equation solving or programming, fostering its broader impact on various scientific fields, such as signal processing, finance, medicine, and biology among a long list.
\end{abstract}


%%  Upper case for every keyword

%\tableofcontents
\end{frontmatter}

\section{Introduction}

In the domain of modern control theory, filtering is an essential subject that has permeated numerous fields, such as signal processing \cite{Candy2016, Roth2017}, weather prediction \cite{Galanis2006, Chen2014}, and aerospace engineering \cite{Ichard2015, Sun2019}. The principal goal of filtering is to attain the ccurate estimation or forecast of a stochastic dynamical system's state, using a set of noisy observations \cite{Jazwinski2007, BainCrisan2009}. For real-world applications, it is crucial that these estimations or forecasts are computed iteratively and in real-time.

Nonlinear filtering, in particular, has a broad spectrum of applications in military, engineering, and commercial industries \cite{Rigatos2011, Rigatos2013, RigatosSiano2011}. The nonlinear filtering problem considered here is to determine estimated states for a given observation history of the following signal-observation model \cite{BainCrisan2009, Jazwinski2007}:

\begin{equation}
\begin{cases}
\mathrm{d} \mathbf{x}(t) = \mathbf{f}(\mathbf{x}(t))\mathrm{d} t + \mathrm{d} \mathbf{v}(t) & \mathbf{x}(0) = x_{0}, \\
\mathrm{d} \mathbf{y}(t) = \mathbf{h}(\mathbf{x}(t))\mathrm{d} t + \mathrm{d} \mathbf{w}(t) & \mathbf{y}(0) = 0,
\end{cases}
\end{equation}
where $\mathbf{x}(t) = (x_1(t), \ldots, x_D(t))^\top \in \mathbb{R}^D$ and $\mathbf{y}(t) = (y_1(t), \ldots, y_M(t))^\top \in \mathbb{R}^M$ are the state and the measurement/observation vectors at time $t$, respectively, $\mathbf{f}(\mathbf{x}) = (f_1(\mathbf{x}), \ldots, f_D(\mathbf{x}))^\top$ and $\mathbf{h}(\mathbf{x}) = (h_1(\mathbf{x}), \ldots, h_M(\mathbf{x}))^\top$ are given vector-valued functions, $\mathbf{v} \in \mathbb{R}^D$ and $\mathbf{w} \in \mathbb{R}^M$ are mutually independent standard Brownian processes. From the main results of Yau and Yau \cite{YauYau1997, YauYau2000, YauYau2008}, the state vector $\mathbf{x}(t)$ can be estimated from the observation vectors $\{\mathbf{y}(s) \,|\, s \in [0, t]\}$ by solving the Kolmogorov equations.

Following the introduction of the famous Kalman filter by Kalman and Bucy in the 1960s \cite{Kalman1960, KalmanBucy1961}, which has been widely applied across various industries, many researchers have focused on studying nonlinear filtering (NLF) theory and developing practical NLF algorithms. A key challenge in NLF is how to find the best estimate of the state from noisy observations. The optimal estimation of the state can be expressed as a minimum mean square error estimate, which is its conditional expectation based on the observation history \cite{Jazwinski2007}.

To achieve the best estimate, one method is to directly approximate the conditional expectation, as seen in the popular extended Kalman filter (EKF) \cite{Jazwinski2007}, unscented Kalman filter (UKF) \cite{JulierUhlmann2004}, and ensemble Kalman filter \cite{Houtekamer1998}. Both EKF and UKF assume that the posterior distribution of the states is essentially Gaussian or nearly Gaussian, which can limit their use.

Another approach to the NLF problem is to calculate the conditional density function of the state. For instance, the particle filter (PF) \cite{Gordon1993} uses the empirical distribution of particles to approximate the posterior density. PF is well-known for its versatility in various NLF problems, although it is not suitable for real-time implementation in high-dimensional systems. In the late 1960s, Duncan, Mortensen, and Zakai independently developed the renowned Duncan-Mortensen-Zakai (DMZ) equation for nonlinear filtering \cite{Duncan1967, Mortensen1994, Zakai1969}, which is satisfied by the unnormalized conditional density function of the states. The DMZ equation, a stochastic differential equation, generally does not have a closed-form solution. For practical use, it is necessary to solve the DMZ equation in real-time and without extensive memory. For finite-dimensional filtering systems, solutions to the DMZ equation can be explicitly built using Lie algebra methods. However, this is only possible for a few types of systems that have finite-dimensional filters \cite{Benes1981, Yau2005}. Since the DMZ equation usually lacks a closed-form solution, many mathematicians have sought good approximations. One such method is the splitting up technique, first described by Bensoussan et al. \cite{Bensoussan1990, Bensoussan1992} and later studied in \cite{Gyongy2003, Nagase1995}.

In the 1990s, Lototsky, Mikulevicius, and Rozovskii provided a recursive, time-based Wiener chaos representation for the optimal nonlinear filter \cite{Lototsky1997}. However, these methods generally require that the drift and observation terms, specifically $f(x)$ and $h(x)$ in system (1), are bounded. Another method to solve the DMZ equation is the direct method, which is typically applicable only to Yau filtering systems where the drift term is a linear function plus a gradient function. This method was introduced in \cite{Yau1994,Yau1994b} and later generalized in \cite{Yau2003, Chen2019}. In 2008, Yau and Yau developed the Yau-Yau algorithm, a new method for solving the "pathwise-robust" DMZ equation in time-invariant systems \cite{YauYau2008}. It has been theoretically proven that the Yau-Yau algorithm can converge to the true solution, as long as the growth rate of the observation $|h|$ is greater than that of the drift $|f|$. Later, Luo and Yau extended the Yau-Yau algorithm to time-varying cases \cite{Luo2013}.

For readers interested in the mathematical principles and computational techniques enabling the software to address high-dimensional nonlinear filtering problems, we direct them to the work of Yueh et al. \cite{Yueh2014,Yueh2014a}. These publications provide a detailed analysis of the underlying mechanisms that power the software's functionality.

In this article, leveraging the powerful capabilities of R and C++, we have crafted an advanced software package. Rooted in the YauYau algorithm, this tool elegantly addresses the complexities of nonlinear filtering problems with remarkable precision and efficiency. This package delivers sophisticated numerical techniques for filtering, while seamlessly integrating an engaging Shiny application designed for dynamic visualization and in-depth exploration. This software package is designed to be concise and user-friendly, enabling users to quickly get started and operate it with ease.

\section{The YauYauAL package}

\subsection{Dependency}

The \texttt{YauYauAL} package (v0.1.0) is built and tested on R version 4.4.2, which can work properly on a standard laptop computer with R version 4.4.2 or higher installed. Also, it is recommended to install RStudio for a better view and interaction with the objects stored in the R environment. The following dependencies are required for data processing: \texttt{Deriv} (v4.1.6), \texttt{RcppArmadillo} (v1.4.2.3-1), \texttt{RcppEigen} (v0.3.4.0.2), and \texttt{Matrix} (v1.7-1). In addition, \texttt{ggplot2} (v3.5.1), \texttt{reshape2} (v1.4.4), \texttt{gridExtra} (v2.3), \texttt{shiny} (v1.10.0), and \texttt{shinythemes} (v1.2.0) are required for visualization.

\subsection{Software Installation Guide}

Begin by installing the prerequisite R package \texttt{devtools}. Open your R console and execute the following command:

\begin{lstlisting}[language=R]
>install.packages("devtools")
\end{lstlisting}

After successful installation, proceed to install \texttt{YauYauAL} directly from the BIMSA-Stat GitHub repository using:

\begin{lstlisting}[language=R]
>devtools::install_github("BIMSA-Stat/YauYauAL")
\end{lstlisting}

For users preferring manual installation, first download the YauYauAL\_0.1.0.tar.gz file from the GitHub repository, and then utilize R's built-in package manager to install the local archive.

Before the \texttt{YauYauAL} can be used in R, it is necessary to import the package using the following command:

\begin{lstlisting}[language=R]
>library(YauYauAL)
\end{lstlisting}

\subsection{Demonstration}

Users can quickly access the interactive graphical user interface of the YauYauAL computational tool using the following R command.

\begin{lstlisting}[language=R]
>YauYauAL::run_app()
\end{lstlisting}

\section{Step-by-step method details}

\subsection{Initialize Parameters}

First, define the dimension of the nonlinear filtering problem and the drift function $\mathbf{f}$ for the state equation, as well as the observation function $\mathbf{h}$ in the observation equation:

\begin{lstlisting}[language=R]
>Dim <- 3
>f <- function(x) {return(c(
  cos(x[1]), 
  cos(x[2]), 
  cos(x[3]))
)}
>h <- function(x) {return(c(
  x[1]^3, 
  x[2]^3, 
  x[3]^3)
)}
\end{lstlisting}

Here, \texttt{Dim} specifies the dimension of the state vector, indicating that the system consists of three state variables. The drift function $\mathbf{f}$ applies the cosine function element-wise to the state vector $\mathbf{x}$, capturing the nonlinear dynamics of the system. Meanwhile, the observation function $\mathbf{h}$ maps each state variable to its cubic power, establishing a highly nonlinear relationship between the states and the observations. This particular configuration is known as the \textbf{Cubic Sensor Problem}, a canonical example in nonlinear filtering that highlights the challenges of state estimation when both the system dynamics and measurement models are nonlinear. The cubic sensor problem is widely recognized for its ability to test the robustness and accuracy of nonlinear filtering algorithms, as traditional linear filtering methods, such as the Kalman Filter, are inadequate due to their reliance on linear assumptions.

\begin{figure}[ht]
\centering
\begin{tikzpicture}[scale=0.85]
% Draw axis
\draw[-] (0,0) node[left] {$0$} -- (10,0) node[right] {$\tma{T}$};

% Place markers
\foreach \x/\l in {
0/{$t^{(1)}_0$}, 
0.5/{$t^{(1)}_1$}, 
2/{$t^{(1)}_{N_t}$}, 
2.5/{$t^{(2)}_{1}$},
4/{$t^{(2)}_{N_t}$}, 
7.7/{$t^{(N_{\tau-1})}_{N_t}$},
8.5/{$t^{(N_\tau)}_{1}$},
10/{$t^{(N_\tau)}_{N_t}$}}
{\draw (\x, 0.1) -- (\x, -0.1) node[below=-0.0cm] {\l};}

\foreach \x/\l in {
1.25/{$\cdots$},
3.25/{$\cdots$},
5.75/{$\cdots$}, 
9.25/{$\cdots$}}
{\node[below=0.2cm] at (\x, 0) {\l};}

\foreach \x/\l in {
5.75/{$\cdots$}}
{\node[above=0.15cm] at (\x, 0) {\l};}

\foreach \x/\l in {
2/{$\rule[-0.5ex]{0.5pt}{1.5ex}\hspace{0.2em}\rule[-0.5ex]{0.5pt}{1.5ex}$}, 
7.7/{$\rule[-0.5ex]{0.5pt}{1.5ex}\hspace{0.2em}\rule[-0.5ex]{0.5pt}{1.5ex}$}}
{\node[below=0.5cm] at (\x, -0.1) {\l};}

\foreach \x/\l in {
2/{$t^{(2)}_{0}$}, 
7.7/{$t^{(N_\tau)}_{0}$}}
{\node[below=0.8cm] at (\x, -0.1)  {\l};}

\foreach \x/\l in {
0/{$\tau_{0}$}, 
2/{$\tau_{1}$}, 
4/{$\tau_{2}$}, 
7.7/{$\tau_{N_{\tau-1}}$},
10/{$\tau_{N_\tau}$}}
{\draw (\x, 0.1) -- (\x, 0.25) node[above] {\l};}

\draw [decorate,decoration={brace,amplitude=5pt,raise=0.7cm}] (0,0) -- (2,0) node[midway,above=0.8cm] {$\Delta \tau$ \tma{(Dtau)}};

\draw [decorate,decoration={brace,amplitude=10pt,raise=1.4cm}] (0,0) -- (10,0) node[midway,above=1.8cm] {$N_\tau$ \tma{(Ntau)}};

\draw [decorate,decoration={brace,mirror,amplitude=5pt,raise=0.75cm}] (0,0) -- (0.5,0) node[midway,above=-1.5cm] {$\Delta t$ \tma{(Dt)}};

\draw [decorate,decoration={brace,mirror,amplitude=7pt,raise=1.5cm}] (0,0) -- (2,0) node[midway,above=-2.3cm] {$N_t$ \tma{(Nt)}};

\draw [decorate,decoration={brace,mirror,amplitude=7pt,raise=2.2cm}] (0,0) -- (10,0) node[midway,above=-3cm] {$N_\tau\cdot N_t$ \tma{(NtauNt)}};

\end{tikzpicture}
\caption{Schematic of Discretized Time Intervals}\label{Td}
\end{figure}
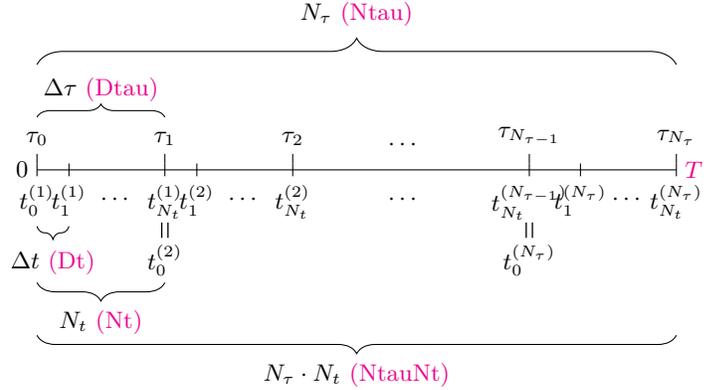

To set up the parameters for our simulation, we proceed as follows:

\begin{lstlisting}
>T <- 5
>Dt <- 0.001
>Dtau <- 5 * Dt                 # 0.005
>Nt <- as.integer(Dtau / Dt)    # 5
>Ntau <- as.integer(T / Dtau)   # 4000
>NtNtau <- as.integer(T / Dt)   # 20000
\end{lstlisting}

In this setup, the terminal time $T$ is set to 20, which defines the total duration of the simulation. The time step $\Delta \tau$ is set to 0.005 for generating data of signals and observations, ensuring that the data is captured at a fine enough resolution to reflect the system's dynamics accurately. The smaller time step $\Delta t$ is designated as 0.001 for solving the Kolmogorov equation, which is crucial for the accuracy of the state estimation in the nonlinear filtering process. The number of time steps $N_t$ and $N_{\tau}$ are calculated based on $T$, $\Delta t$, and $\Delta \tau$ respectively, determining how often the system state is updated and observations are recorded throughout the simulation. These parameters are essential for configuring the simulation environment in the \texttt{YauYauAL} R package, allowing us to effectively address the nonlinear filtering problem. Figure \ref{Td} illustrates this process.

\subsection{State and Observation Generator}

The simulation of state and observation sequences is conducted using the following R script:

\begin{lstlisting}[language=R]
>seed_value <- 42
>result <- Simulate_State_Obser(
  Dt = 0.001, 
  Ntau = 4000, 
  NtNtau = 20000, 
  f = f, 
  h = h, 
  Dim = 3, 
  seed = 42)
>x <- result$x
>y <- result$y
\end{lstlisting}

In this simulation, a seed value of 42 is set for the random number generator to ensure the reproducibility of results. The function \texttt{Simulate\_State\_Obser} is employed with parameters $\Delta t = 0.001$, $N_{\tau} = 4000$, and $N_{\tau}\cdot N_{t} = 20000$ to define the temporal resolution and duration of the simulation. The nonlinear dynamics are modeled through the drift function $f$ and the observation function $h$, with the state dimension set to 3. The sequences $\mathbf{x}$ and $\mathbf{y}$, representing the state and observations, are extracted from the result object for subsequent analysis with nonlinear filtering algorithms.

\subsection{Discretization}

The discretization process involves creating a grid of points in the state space to facilitate the numerical solution of the Kolmogorov equation:

\begin{lstlisting}[language=R]
>Ds <- 0.5
>s <- seq(min(x), max(x)+Ds, by = Ds)
>Ns <- length(s)
>s <- ExpandGrid(Dim,s)
\end{lstlisting}

In the above code, we first define the grid spacing $\Delta s$. Then, we generate a sequence of grid points $s$ that spans from the minimum value of the state sequence $x$ to slightly beyond its maximum value, incremented by $\Delta s$. The number of grid points $N_{s}$ is determined by the length of this sequence. Finally, the \texttt{ExpandGrid} function is used to create a multi-dimensional grid covering the entire state space based on the state vector dimension \texttt{Dim}. This grid will serve as the basis for discretizing the state space in subsequent filtering computations.

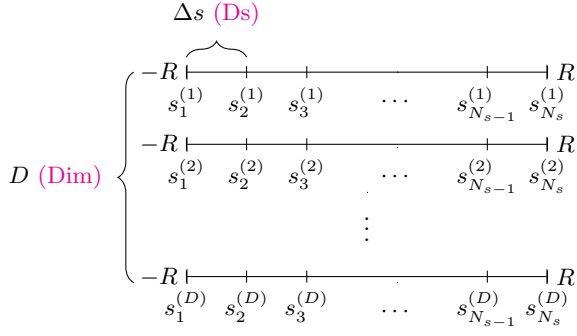
\begin{figure}[ht]
    \centering
    \begin{tikzpicture}[scale=0.8]
    \draw[|-|] (0,0) node[left] {$-R$} -- (6,0) node[right] {$R$};
    \foreach \x/\l in {
    0/{$s^{(1)}_1$}, 
    1/{$s^{(1)}_2$}, 
    2/{$s^{(1)}_3$}, 
    5/{$s^{(1)}_{N_{s-1}}$}, 
    6/{$s^{(1)}_{N_s}$}}
    {\draw (\x, 0.1) -- (\x, -0.1) node[below=-0.0cm] {\l};}
    \foreach \x/\l in {
    3.5/{$\cdots$}}
    {\draw (\x, 0.0) -- (\x, -0.0) node[below=0.2cm] {\l};}
    \draw [decorate,decoration={brace,amplitude=6pt,raise=0.2cm}] (0,0) -- (1,0) node[midway,above=0.5cm] {$\Delta s$ \tma{(Ds)}};

    \draw [decorate,decoration={brace,mirror,amplitude=6pt,raise=0.7cm}] (0,0) -- (0,-3.5) node[midway,left=1cm] {$D$ \tma{(Dim)}};

    \draw[|-|] (0,-1.2) node[left] {$-R$} -- (6,-1.2) node[right] {$R$};
    \foreach \x/\l in {
    0/{$s^{(2)}_1$}, 
    1/{$s^{(2)}_2$}, 
    2/{$s^{(2)}_3$}, 
    5/{$s^{(2)}_{N_{s-1}}$}, 
    6/{$s^{(2)}_{N_s}$}}
    {\draw (\x, -1.1) -- (\x, -1.3) node[below=-0.0cm] {\l};}
    \foreach \x/\l in {
    3.5/{$\cdots$}}
    {\draw (\x, -1.25) -- (\x, -1.25) node[below=0.2cm] {\l};}

    \foreach \x/\l in {
    3/{$\vdots$}}
    {\draw (\x, -2) -- (\x, -2) node[below=0.0cm] {\l};}

    \draw[|-|] (0,-3.4) node[left] {$-R$} -- (6,-3.4) node[right] {$R$};
    \foreach \x/\l in {
    0/{$s^{(D)}_1$}, 
    1/{$s^{(D)}_2$}, 
    2/{$s^{(D)}_3$}, 
    5/{$s^{(D)}_{N_{s-1}}$}, 
    6/{$s^{(D)}_{N_s}$}}
    {\draw (\x, -3.3) -- (\x, -3.5) node[below=-0.0cm] {\l};}
    \foreach \x/\l in {
    3.5/{$\cdots$}}
    {\draw (\x, -3.5) -- (\x, -3.5) node[below=0.2cm] {\l};}
    
    \end{tikzpicture}
    \caption{Schematic Representation of Spatial Discretization Across Dimensions}\label{Sd}
\end{figure}

Figure \ref{Sd} illustrates the process of spatial discretization.

Next, we generate the necessary matrices and parameters for the discretization process:

\begin{lstlisting}[language=R]
>D <- generateD(Dim,Ns,Ds)
>Lambda <- computeLambda(Dim,Ns,Dt,Ds)
>df <- generate_derivative(f)
>B <- computeB(s,D,Dt,Ds,f,df,h)
\end{lstlisting}

In the above code, the function \texttt{generateD} creates the matrix $D$ which represents the discrete derivative operator based on the state dimension \texttt{Dim}, the number of grid points $N_{s}$, and the grid spacing $\Delta s$. The function \texttt{computeLambda} calculates the matrix $\Lambda$ using the state dimension \texttt{Dim}, the number of grid points $N_{s}$, the time step $\Delta t$, and the grid spacing $\Delta s$. The function \texttt{generate\_derivative} generates the derivative function $df$  based on the drift function $f$. Finally, the function \texttt{computeB} computes the matrix $B$ using the grid points \texttt{s}, the matrix $D$, the time step $\Delta t$, the grid spacing $\Delta s$, the drift function $f$, its derivative $df$, and the observation function $h$. These matrices and parameters are essential for the subsequent steps in the nonlinear filtering process, particularly for solving the Kolmogorov equation numerically.

\subsection{State Estimation}

The state estimation process involves computing the posterior distribution of the state variables given the observations. This is achieved using the following R script:

\begin{lstlisting}[language=R]
>Iu <- wrap_outiu_function(s, NtNtau, Ntau, Nt, Dim, y, h, Lambda, B, Ns, NormalizedExp, DST_Solver)
\end{lstlisting}

The function computes the posterior distribution of the state variables at each time step using the grid points, time steps, state dimension, observations, observation function, noise parameters, grid size, normalization function, and solver function. The Discrete Sine Transform to efficiently solve the Kolmogorov equation for accurate state estimation in nonlinear systems.

\subsection{Interactive Implementation of the Algorithm with Shiny}

\begin{figure*}[t]
\centering
\includegraphics[width=1.0\linewidth]{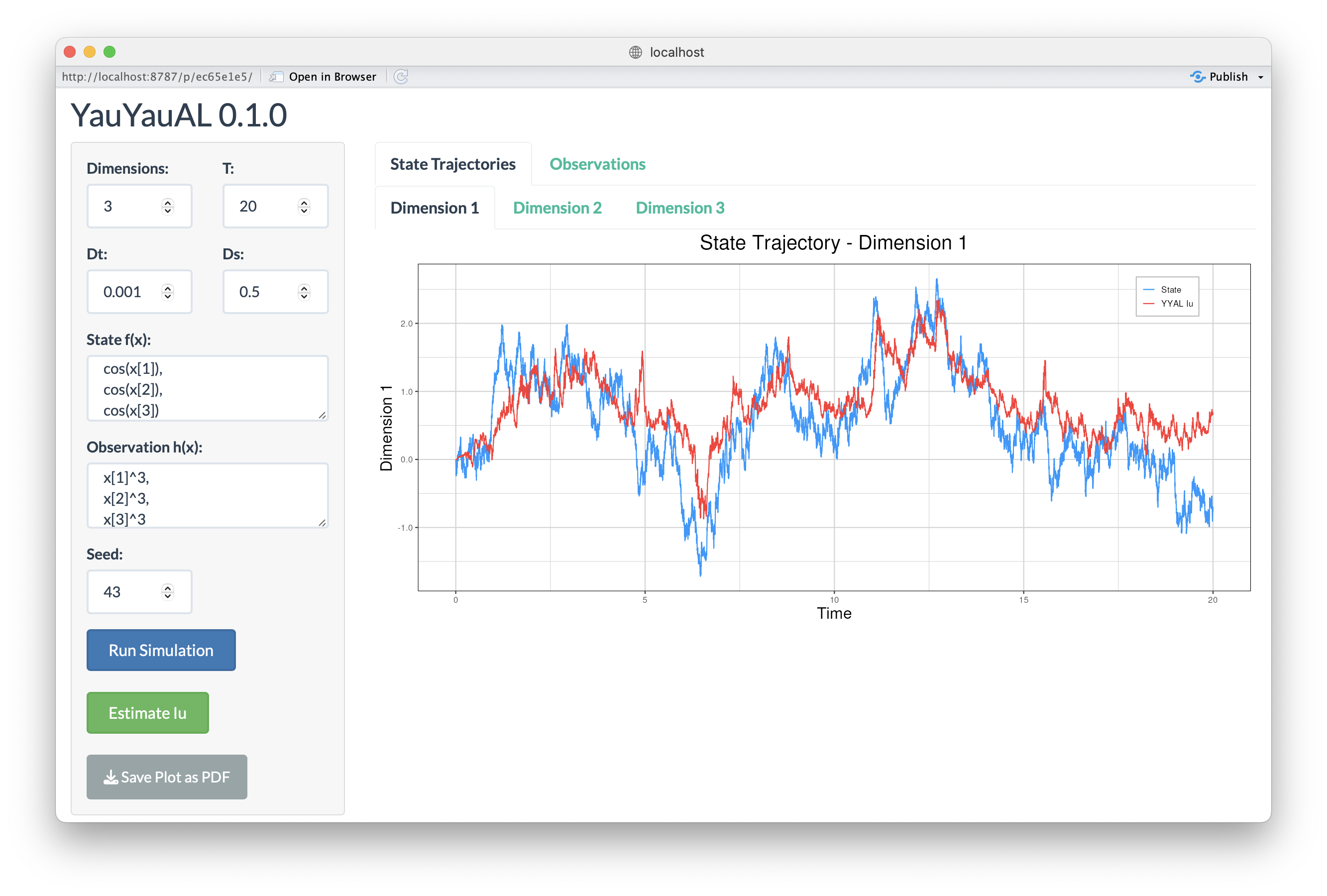}
\caption{The interface of the interactive YauYauAL software based on Shiny.}\label{shiny}
\end{figure*}

You can launch the homepage of the interactive software interface using the following R command:

\begin{lstlisting}[language=R]
>YauYauAL::run_app()
\end{lstlisting}

Figure~\ref{shiny} shows the interface of our interactive software. When defining custom filters, the algorithm may encounter significant delays in execution. In such scenarios, Shiny might not update the results in real-time. We recommend that users run the scripts outlined in subsections 3.1–3.4 independently to obtain the results, followed by visualization using the plot function.

\section{Numerical Experiments}

\subsection{almost linear sensor problem}

\textbf{Example 4.1.~}
The almost linear sensor problem is as follows:
\begin{equation}\label{exm1}
\begin{cases}
\mathrm{d} x(t) = d v(t)\\
\mathrm{d} y(t) = x(t)\left(1+0.25\cos{x(t)}\right)\mathrm{d} t + \mathrm{d} w(t)
\end{cases}   
\end{equation}
where $x(t)$, $y(t) \in \mathbb{R}$, $v(t)$, $w(t)$ are independent standard scalar Brownian motion processes. The parameters $\mathrm{Dim}=3$, $T=50$, $\Delta t = 0.0001$, $\Delta \tau = 0.0005$, $\Delta s = 0.5$.

\begin{figure}[H]
\small
\centering
\includegraphics[width=0.95\linewidth]{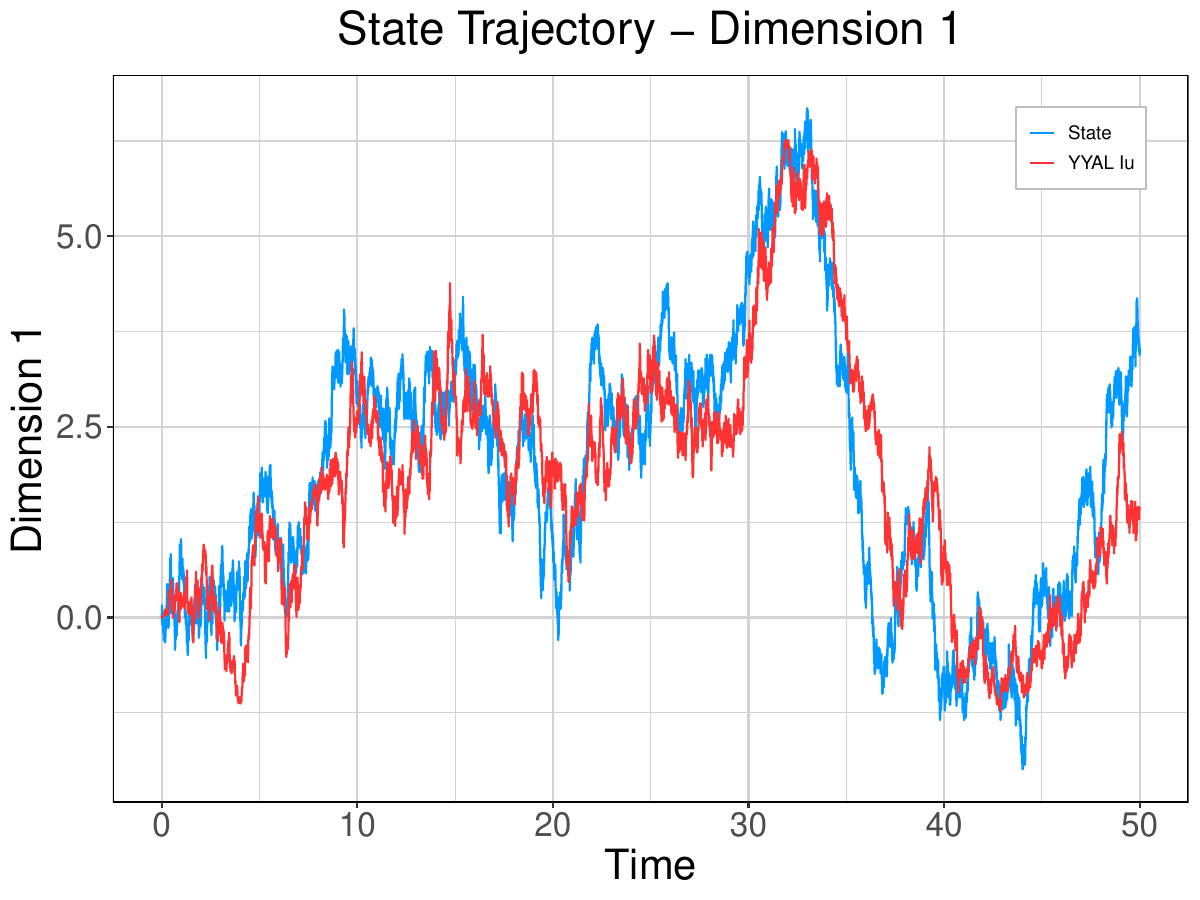}
\caption{Estimations of the almost linear sensor for the model Eq.\ref{exm1}, and  $T=50$, with the time step $\Delta t = 0.0001$. Blue: real state; Red: Yau-Yau filter.}\label{fig-exm1}
\end{figure}

Figure~\ref{fig-exm1} illustrates the results of the Yau-Yau Filter when dealing with the almost linear sensor problem.

\subsection{cubic sensor problem}

\textbf{Example 4.2.~}

The cubic sensor problem \cite{YauYau2006} is as follows:
\begin{equation}\label{exm2}
\begin{cases}
\mathrm{d} x(t) = \cos{x(t)}\mathrm{d} t + \mathrm{d} v(t)\\
\mathrm{d} y(t) = x^{3}\mathrm{d} t + \mathrm{d} w(t)
\end{cases}   
\end{equation}
where $x(t)$, $y(t) \in \mathbb{R}$, $v(t)$, $w(t)$ are independent standard scalar Brownian motion processes. The parameters $\mathrm{Dim}=3$,$T=20$, $\Delta t = 0.001$, $\Delta \tau = 0.005$, $\Delta s = 0.5$.

\begin{figure*}[htpb]
\centering
\includegraphics[width=0.3\linewidth]{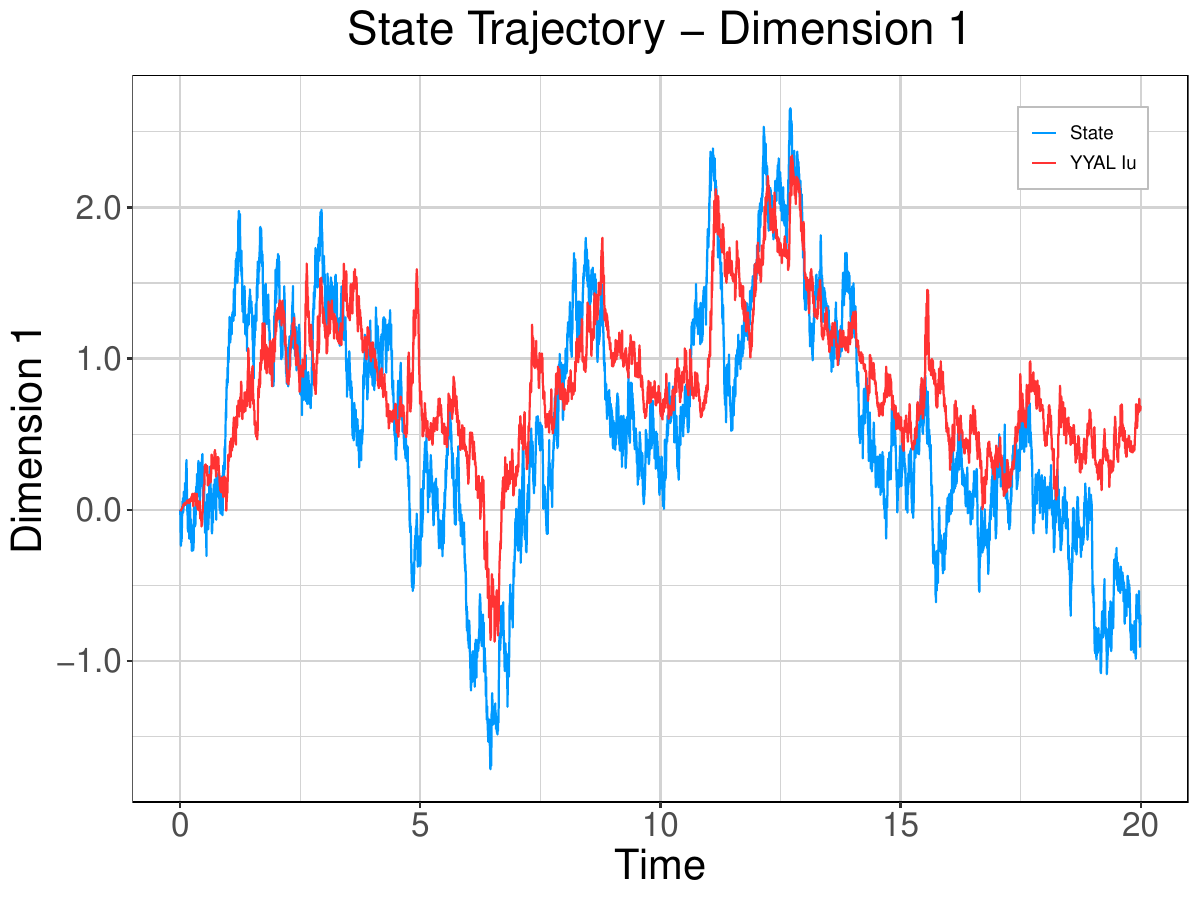}
\includegraphics[width=0.3\linewidth]{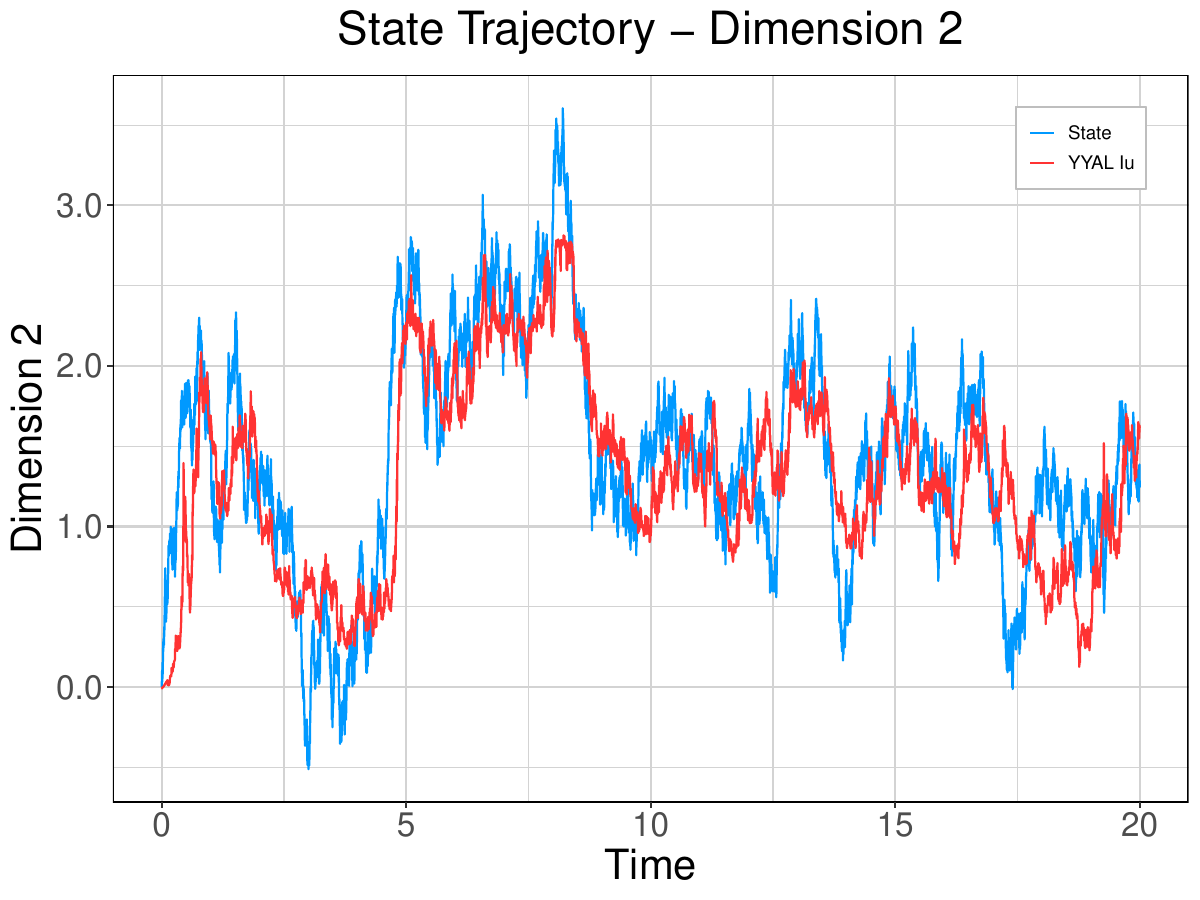}
\includegraphics[width=0.3\linewidth]{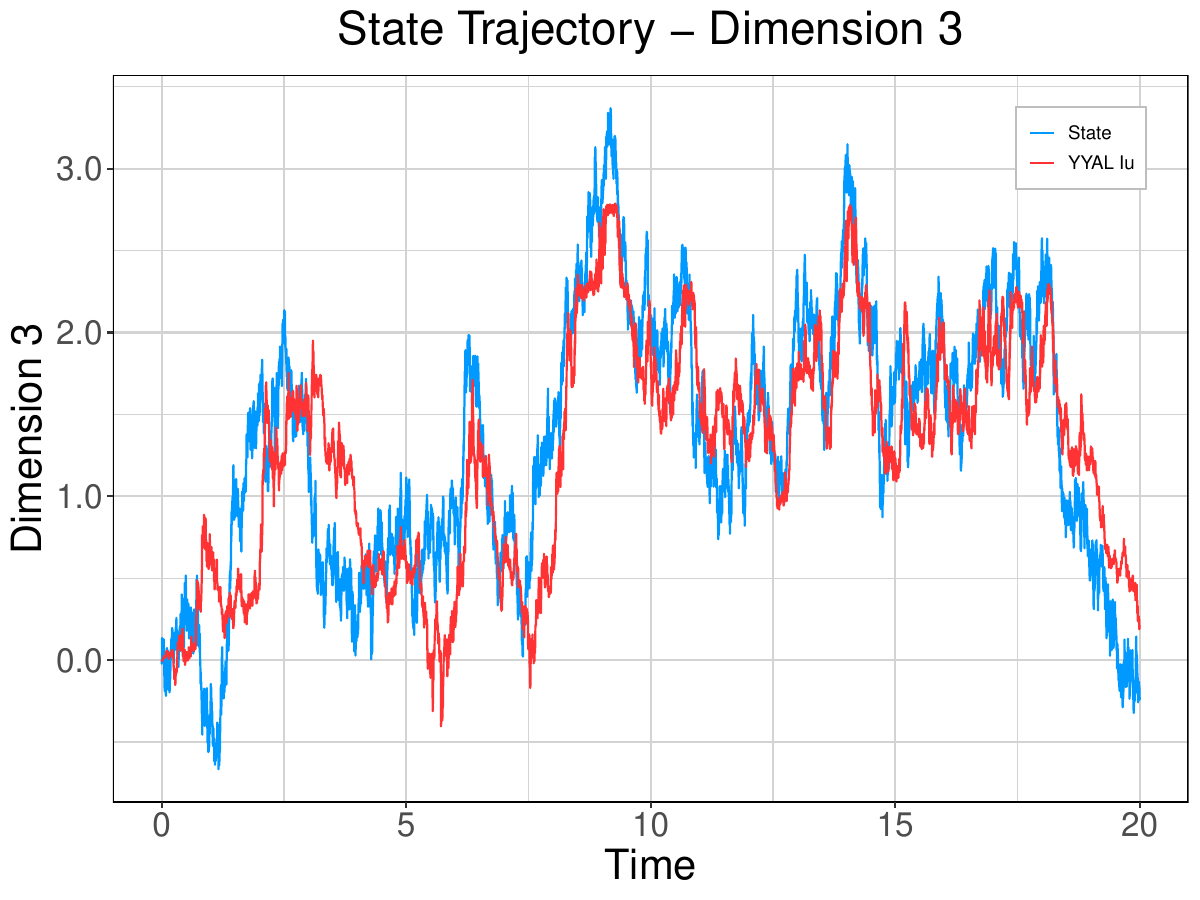}
\caption{Estimations of the 3-D Cubic sensor for the model Eq.\ref{exm2}, and  $T=20$, 
with the time step $\Delta t = 0.001$.
Blue: real state; 
Red: Yau-Yau filter.}\label{fig-exm2}
\end{figure*}

Figure~\ref{fig-exm2} presents the outcomes of the Yau-Yau Filter applied to the 3D cubic sensor problem.

\section{Discuss}

The development of the Yau-YauAL software package represents a significant step forward in the practical application of the Yau-Yau nonlinear filter. By integrating the powerful capabilities of R with the computational efficiency of C++ through Rcpp, Yau-YauAL effectively addresses the challenges associated with implementing complex filtering algorithms. The use of finite difference methods to solve the Kolmogorov forward equation ensures that the tool can handle a wide range of complex systems while maintaining stability and accuracy in its solutions.

The inclusion of an intuitive Shiny-based interface is particularly noteworthy. This feature democratizes access to advanced filtering techniques by allowing users to adjust parameters and visualize results in real-time, without the need for extensive programming knowledge. This is especially beneficial for researchers in fields such as finance, signal processing, and biology, where the application of nonlinear filtering can yield valuable insights but may be hindered by the complexity of the underlying mathematics.

The modular design and open-source nature of Yau-YauAL further enhance its utility. By allowing for customization and community-driven development, the software can be adapted to meet the specific needs of various users and evolve with advancements in the field. This collaborative approach not only fosters innovation but also ensures that the tool remains relevant and effective in a rapidly changing scientific landscape.

However, it is important to acknowledge potential limitations. While Yau-YauAL has been designed to be user-friendly, the complexity of some nonlinear filtering problems may still require a certain level of expertise to fully exploit the tool's capabilities. Additionally, the performance of the finite difference methods employed may be influenced by the specific characteristics of the system being analyzed, and further research may be needed to optimize these methods for different applications.

Future work could focus on expanding the range of numerical methods available within Yau-YauAL to provide users with even more options for solving nonlinear filtering problems. Additionally, incorporating machine learning techniques could enhance the tool's ability to handle large datasets and improve the accuracy of its solutions. Collaboration with researchers across disciplines will be crucial in identifying new applications and refining the software to meet emerging challenges.

In conclusion, Yau-YauAL has the potential to significantly impact the field of nonlinear filtering by making advanced techniques more accessible and user-friendly. Its innovative design and open-source framework position it as a valuable resource for both current and future interdisciplinary research endeavors.

% \clearpage
\section*{Code and data availability}
The code and data related to this study are publicly available on GitHub. The repository can be accessed via the following link: \url{https://github.com/BIMSA-Stat/YauYauAL}. This repository contains the complete implementation of the YauYauAL software package, including the source code, documentation, and example datasets used in this research. Users are encouraged to explore the repository for further details and to utilize the provided resources for their own research and development purposes.

\section*{Acknowledgements}
We thank the colleagues at the Beijing Key Laboratory of Topological Statistics and Applications for Complex Systems at the Beijing Institute of Mathematical Sciences and Applications for their contributions to this work. The authors thank the anonymous referees for their careful reading and valuable suggestions. This work was supported in part by the Natural Science Foundation of Inner Mongolia (No. 2024MS01017) and the First Class Discipline Project, Inner Mongolia Autonomous Region, China (No. YLXKZX-NSD-007)


\begin{thebibliography}{9}
%%  Put author names in \textsc{} command in order to use small caps font


% ========== A ==========

% ========== B ==========

\bibitem{BainCrisan2009}
\textsc{Bain, A.} and \textsc{Crisan, D.} (2009). \textit{Fundamentals of Stochastic Filtering}. Springer Science Business Media, 1st ed.
\MR{2454694}

\bibitem{Benes1981}
\textsc{Benes, V. E.} (1981). Exact finite-dimensional filters for certain diffusions with nonlinear drift. \textit{Stochastics} \textbf{5}(1--2):65--92.

\bibitem{Bensoussan1990}
\textsc{Bensoussan, A.}, \textsc{Glowinski, R.}, and \textsc{Rascanu, A.} (1990). Approximation of the Zakai equation by the splitting up method. \textit{SIAM J. Control Optim.} \textbf{28}(6):1420--1431.

\bibitem{Bensoussan1992}
\textsc{Bensoussan, A.}, \textsc{Glowinski, R.}, and \textsc{Rascanu, A.} (1992). Approximation of some stochastic differential equations by the splitting up method. \textit{Appl. Math. Optim.} \textbf{25}(1):81--106.

% ========== C ==========

\bibitem{Candy2016}
\textsc{Candy, J. V.} (2016). 
\textit{Bayesian Signal Processing: Classical, Modern, and Particle Filtering Methods}, 1st ed. John Wiley \& Sons, Hoboken, NJ.

\bibitem{Chen2014}
\textsc{Chen, K.} and \textsc{Yu, J.} (2014). Short-term wind speed prediction using an unscented Kalman filter based state-space support vector regression approach. \textit{Applied Energy} \textbf{113}:690--705.

\bibitem{Chen2019}
\textsc{Chen, X.}, \textsc{Shi, J.}, and \textsc{Yau, S. S.-T.} (2019). Real-time solution of time-varying Yau filtering problems via direct method and Gaussian approximation. \textit{IEEE Trans. Autom. Control} \textbf{64}(4):1648--1654.

% ========== D ==========

\bibitem{Duncan1967}
\textsc{Duncan, T. E.} (1967). Probability densities for diffusion processes with applications to nonlinear filtering theory and detection theory. Stanford California, Stanford, CA, USA, Tech. Rep. TR-7001-4.

% ========== E ==========
% ========== F ==========
% ========== G ==========

\bibitem{Galanis2006}
\textsc{Galanis, G.}, \textsc{Louka, P.}, \textsc{Katsafados, P.}, \textsc{Pytharoulis, I.}, and \textsc{Kallos, G.} (2006). Applications of Kalman filters based on non-linear functions to numerical weather predictions. \textit{Annales Geophysicae} \textbf{24}(1):2451--2460.

\bibitem{Gordon1993}
\textsc{Gordon, N. J.}, \textsc{Salmond, D. J.}, and \textsc{Smith, A. F. M.} (1993). Novel approach to nonlinear/non-Gaussian Bayesian state estimation. \textit{IEEE Proc. F, Radar Signal Process.} \textbf{140}(2):107--113.

\bibitem{Gyongy2003}
\textsc{Gyongy, I.} and \textsc{Krylov, N.} (2003). On the splitting-up method and stochastic partial differential equations. \textit{Ann. Probab.} \textbf{31}(2):564--591.

% ========== H ==========

\bibitem{Houtekamer1998}
\textsc{Houtekamer, P. L.} and \textsc{Mitchell, H. L.} (1998). Data assimilation using an ensemble Kalman filter technique. \textit{Monthly Weather Rev.} \textbf{126}(3):796--811.

% ========== I ==========

\bibitem{Ichard2015}
\textsc{Ichard, C.} (2015). \textit{Random Media and Processes Estimation Using Non-linear Filtering Techniques: Application to Ensemble Weather Forecast and Aircraft Trajectories}. PhD thesis, Université de Toulouse, Université Toulouse III--Paul Sabatier.

\bibitem{Ito1996}
\textsc{Ito, K.} (1996). Approximation of the Zakai equation for nonlinear filtering. \textit{SIAM J. Control Optim.} \textbf{34}(2):620--634.

\bibitem{Ito2000}
\textsc{Ito, K.} and \textsc{Rozovskii, B.} (2000). Approximation of the Kushner equation for nonlinear filtering. \textit{SIAM J. Control Optim.} \textbf{38}(3):893--915.

% ========== J ==========

\bibitem{Jazwinski2007}
\textsc{Jazwinski, A. H.} (2007). \textit{Stochastic Processes and Filtering Theory}. Courier Corporation.

\bibitem{JulierUhlmann2004}
\textsc{Julier, S. J.} and \textsc{Uhlmann, J. K.} (2004). Unscented filtering and nonlinear estimation. \textit{Proc. IEEE} \textbf{92}(3):401--422.

% ========== K ==========

\bibitem{Kalman1960}
\textsc{Kalman, R. E.} (1960). A new approach to linear filtering and prediction problems. \textit{J. Basic Eng.} \textbf{82}:35--45.

\bibitem{KalmanBucy1961}
\textsc{Kalman, R. E.} and \textsc{Bucy, R. S.} (1961). New results in linear filtering and prediction theory. \textit{J. Basic Eng.} \textbf{83}(1):95--108.

% ========== L ==========

\bibitem{Lototsky1997}
\textsc{Lototsky, S.}, \textsc{Mikulevicius, R.}, and \textsc{Rozovskii, B. L.} (1997). Nonlinear filtering revisited: A spectral approach. \textit{SIAM J. Control Optim.} \textbf{35}(2):435--461.

\bibitem{Luo2013}
\textsc{Luo, X.} and \textsc{Yau, S. S.-T.} (2013). Hermite spectral method to 1-D forward Kolmogorov equation and its application to nonlinear filtering problems. \textit{IEEE Trans. Autom. Control} \textbf{58}(10):2495--2507.

% ========== M ==========

\bibitem{Mortensen1994}
\textsc{Mortensen, R. E.} (1966). Optimal control of continuous time stochastic systems. Ph.D. dissertation, Electron. Res. Lab., California Univ. Berkeley, Berkeley, CA, USA. 

\bibitem{Yueh2014}
\textsc{Yueh, M.-H.}, \textsc{Lin, W.-W.}, and \textsc{Yau, S.-T.} (2014).
An efficient numerical method for solving high-dimensional nonlinear filtering problems. \textit{Communications in Information and Systems} \textbf{14}(4):243--262.

\bibitem{Yueh2014a}
\textsc{Yueh, M.-H.}, \textsc{Lin, W.-W.}, and \textsc{Yau, S.-T.} (2014).
An Efficient Algorithm of Yau-Yau Method for Solving Nonlinear Filtering Problems. \textit{Communications in Information and Systems} \textbf{14}(2):111--134.

% ========== N ==========

\bibitem{Nagase1995}
\textsc{Nagase, N.} (1995). Remarks on nonlinear stochastic partial differential equations: An application of the splitting-up method. \textit{SIAM J. Control Optim.} \textbf{33}(6):1716--1730.

% ========== O ==========
% ========== P ==========
% ========== Q ==========
% ========== R ==========

\bibitem{Rigatos2011}
\textsc{Rigatos, G. G.} (2011). \textit{Modelling and Control for Intelligent Industrial Systems: Adaptive Algorithms in Robotics and Industrial Engineering}. Springer-Verlag Berlin Heidelberg.

\bibitem{RigatosSiano2011}
\textsc{Rigatos, G. G.} and \textsc{Siano, P.} (2011). Sensorless control of electric motors with Kalman filters: applications to robotic and industrial systems. \textit{International Journal of Advanced Robotic Systems} \textbf{8}(6):62--80.

\bibitem{Rigatos2013}
\textsc{Rigatos, G. G.} (2013). \textit{Nonlinear Estimation and Applications to Industrial Systems Control}. Nova Science Publishers Inc.

\bibitem{Roth2017}
\textsc{Roth, M.}, \textsc{Hendeby, G.}, \textsc{Fritsche, C.}, and \textsc{Gustafsson, F.} (2017). The ensemble Kalman filter: a signal processing perspective. \textit{EURASIP Journal on Advances in Signal Processing} \textbf{2017}(1):1--16.

% ========== S ==========

\bibitem{Shi2018}
\textsc{Shi, J.}, \textsc{Yang, Z.}, and \textsc{Yau, S. S. T.} (2018). Direct method for Yau filtering system with nonlinear observations. \textit{Int. J. Control} \textbf{91}(3):678--687.

\bibitem{Sun2019}
\textsc{Sun, J.}, \textsc{Blom, H. A.}, \textsc{Ellerbroek, J.}, and \textsc{Hoekstra, J. M.} (2019). Particle filter for aircraft mass estimation and uncertainty modeling. \textit{Transportation Research Part C: Emerging Technologies} \textbf{105}:145--162.

% ========== T ==========
% ========== U ==========
% ========== V ==========
% ========== W ==========

\bibitem{Wong1983}
\textsc{Wong, W. S.} (1983). New classes of finite-dimensional nonlinear filters. \textit{Syst. Control Lett.} \textbf{3}(3):155--164.

% ========== X ==========
% ========== Y ==========

\bibitem{Yau1994}
\textsc{Yau, S. S.-T.} and \textsc{Yau, S. T.} (1994). New direct method for Kalman-Bucy filtering system with arbitrary initial condition. In \textit{Proc. 33rd IEEE Conf. Decis. Control}, vol. 2, pp. 1221--1225.

\bibitem{Yau1994b}
\textsc{Yau, S. S.-T.} (1994). Finite-dimensional filters with nonlinear drift. I: A class of filters including both Kalman-Bucy and Bene's filters. \textit{J. Math. Syst. Estimation Control} \textbf{4}(2):181--203.

\bibitem{YauYau1997}
\textsc{Yau, S. T.} and \textsc{Yau, S. S.-T.} (1997). Finite dimensional filters with nonlinear drift iii: Duncan-Mortensen-Zakai equation with arbitrary initial condition for kalman-bucy filtering system and benes filtering system. \textit{IEEE Transactions on Aerospace and Electronic Systems} \textbf{33}:1277--1294.

\bibitem{YauYau2000}
\textsc{Yau, S.-T.} and \textsc{Yau, S. S.-T.} (2000). Real time solution of nonlinear filtering problem without memory I. \textit{Mathematical Research Letters} \textbf{7}:671--693.
\MR{1809293}

\bibitem{Yau2003}
\textsc{Yau, S. S.-T.} and \textsc{Lai, Y.-T.} (2003). Explicit solution of DMZ equation in nonlinear filtering via solution of ODEs. \textit{IEEE Trans. Autom. Control} \textbf{48}(3):505--508.

\bibitem{Yau2004}
\textsc{Yau, S. T.} and \textsc{Yau, S. S.-T.} (2004). Nonlinear filtering and time varying Schrödinger equation. \textit{IEEE Trans. Aerosp. Electron. Syst.} \textbf{40}:284--292.

\bibitem{Yau2004b}
\textsc{Yau, S. S.-T.}, \textsc{Yan, C.}, and \textsc{Yau, S.-T.} (2004). Linear filtering with nonlinear observations. In \textit{Proc. 43rd IEEE Conf. Decis. Control (CDC)}, vol. 2, pp. 2112--2117.

\bibitem{Yau2005}
\textsc{Yau, S. S.-T.} and \textsc{Hu, G.-Q.} (2005). Classification of finite-dimensional estimation algebras of maximal rank with arbitrary state-space dimension and Mitter conjecture. \textit{Int. J. Control} \textbf{78}(10):689--705.

\bibitem{YauYau2006}
\textsc{Yan, C.} and \textsc{Yau, S. S.-T.} (2006). A new suboptimal filter and numerical solutions for the cubic sensor problem. In \textit{2006 IEEE International Conference on Networking, Sensing and Control}, pages 351--356.

\bibitem{YauYau2008}
\textsc{Yau, S.-T.} and \textsc{Yau, S. S.-T.} (2008). Real time solution of nonlinear filtering problem without memory II. \textit{SIAM J. Control Optim.} \textbf{47}(1):163--195.
\MR{2373467}

% ========== Z ==========

\bibitem{Zakai1969}
\textsc{Zakai, M.} (1969). On the optimal filtering of diffusion processes. \textit{Zeitschrift für Wahrscheinlichkeitstheorie und verwandte Gebiete} \textbf{11}(3):230--243.

\end{thebibliography}
\end{document}